# GSC2.3 N152008120 - a new SU UMa-type dwarf nova in Draco

David Boyd, Keith Graham, Taichi Kato, Robert Koff, Ian Miller, Arto Oksanen, Roger Pickard, Gary Poyner

**Abstract**

We report observations during 2008 October of the first recorded superoutburst of a previously unknown SU UMa-type dwarf nova in Draco located at 19h 14m 43.52s +60° 52' 14.1" (J2000). Simbad lists a 21st magnitude star at this position with identifiers GSC2.3 N152008120 and USNO-B1.0 1508-0249096. The outburst reached magnitude 14.9, its amplitude was approximately 6 magnitudes and its duration at least 11 days. About 11 days after the end of the main outburst there was a short-lived rebrightening by more than 2 magnitudes. Superhumps were observed with a mean period of 0.07117(1) d and amplitude 0.12 mag. There was a distinct shortening in the superhump period around cycle 80 with $P_{sh}$ = 0.07137(2) d before and $P_{sh}$ = 0.07091(2) d after. We saw weak evidence of an increasing $P_{sh}$ before cycle 80 with $dP_{sh}/dt = 3.4(2.0) * 10^{-5}$.

**Discovery and early outburst observations**

A new optical transient source in Draco was detected by K Itagaki (Yamagata, Japan) on 2008 October 2.533 UT on a CCD survey image taken with a 0.21-m reflector. An image taken on October 2.539 UT with a 0.60-m reflector showed the object at about magnitude 14.9 with position 19h 14m 43s.55 +60° 52' 13".9 (J2000). S Nakano (Sumoto, Japan) recorded a magnitude of 15.0 on October 3.387. H Maehara observed magnitudes of V=15.16 on October 3.564 UT and B=15.07 on October 3.566 UT indicating the object was very blue, as expected for an outbursting dwarf nova. Maehara also performed 3 hours of time series photometry which showed superhump-like modulations with an amplitude of 0.2 magnitude. These observations were reported to the Central Bureau for Astronomical Telegrams (CBAT) by H. Yamaoka of Kyushu University and the object was posted on the CBAT Unconfirmed Observations webpage [1].

These early observations indicated that this was probably a previously unknown dwarf nova seen in outburst for the first time. A report of Maehara's observations was distributed by VSNET on October 6 [2] which alerted observers worldwide to this new object.

**Previous observations**

CBAT issued Electronic Telegram No 1535 on October 10 which reported that nothing was visible at the position of the object on Itagaki's survey image taken on June 12.597 (limiting magnitude 17.5) or on the Digitized Sky Survey from 1991 August 30 (F) and 1992 June 28 (J) to limiting magnitude about 21.

A search on Simbad [3] at the reported position reveals an object with identifiers GSC2.3 N152008120 and USNO-B1.0 1508-0249096. The USNO B1.0 catalogue gives the position 19h 14m 43.39s +60° 52' 14.4" (J2000) and magnitudes B=21.1 and R=19.8. A search in VizieR [4] finds no X-ray source or suspected variable at that position. The object has also been referred to as OT J191443.6+605214 and Var Dra 2008.

**Our observations**

Over the following 3 weeks the authors obtained 44 hours of time series CCD photometry comprising 1653 measurements which initially showed well-developed superhumps. This confirmed the identification of this object as a SU UMa-type dwarf nova. Our observations recorded its decline from outburst towards quiescence punctuated by a single rebrightening. A log of these time series observations is given in Table 1 and the equipment used is listed in Table 2. An image of the dwarf nova taken by Oksanen on October 8 is shown in Figure 1.

**Photometry**

As the dwarf nova was relatively faint, most observations were obtained with a clear filter or, in the case of one observer, a luminance filter. All images were dark-subtracted and flat-fielded and instrumental magnitudes obtained using aperture photometry. Comparison star magnitudes were derived from the USNO-A2.0 catalogue using Kidger's formulae [5] to convert from USNO B and R magnitudes to Landolt V and B-V. Given the blue colour of the object, we selected 5 stars which had B-V values between 0.3 and 0.5. These comparison stars are listed in Table 3 and were used to measure all our images.

Because of its blue colour, the different blue sensitivity of the CCD cameras used by each observer and the different filters used, there was some variation in the magnitudes obtained for the dwarf nova by different observers at or around the same time, even though a consistent set of comparison stars was used. By experiment we found the most satisfactory way of bringing the magnitude measurements of different observers into alignment was by using the fitted maximum magnitudes of superhumps which were obtained from the fits to the superhump profiles described below. The resulting combined light curve for the outburst is shown in Figure 2. In this plot, time series data from the authors are marked as solid dots without error bars, mean values measured during the decline as solid dots with error bars, and single observations as open dots including the early outburst observations from various sources mentioned above.

As no magnitude estimates were obtained on the rise, we cannot say when maximum occurred. From the observations in Figure 2, the outburst lasted at least 11 days from JD 2454742 to JD 2454753 fading at an average rate of 0.14 mag/day during this period. It then entering a period of more rapid decline at 0.31 mag/day. During this decline, an image taken by Poyner with the Bradford Robotic Telescope (BRT) on JD 2454764 clearly shows the dwarf nova at magnitude 17.0, more than 2 magnitudes brighter than expected. This late rebrightening lasted less than 5 days before a return to the previous decline trajectory. Given historical observations in quiescence and our final measurement, we can say that the outburst amplitude was approximately 6 magnitudes.

**Astrometry**

Astrometry of the dwarf nova on five images taken under good conditions on 2008 October 8 using Astrometrica [6] and the USNO-B1.0 catalogue gives a mean position of 19h 14m 43.52s +60° 52' 14.1" (J2000) with an estimated uncertainty of 0.13".

**Superhump maximum timing analysis**

Figure 3 shows superhumps recorded on JD 2454748. These have an amplitude of approximately 0.2 magnitude. During the outburst, 13 superhumps were sufficiently well recorded by European and American observers to provide measurements of the times of superhump maximum. These times were obtained by a quadratic fit to the superhump profile around each maximum. The fits also gave magnitudes at superhump maximum which were used to align the observations of different observers in magnitude as described earlier. Timings for 24 superhumps obtained by Japanese observers were subsequently provided by Kato. These timings were obtained by fitting a standard superhump template to each superhump. Comparing superhump timings obtained from the same data using both methods revealed a small systematic difference which was compensated for by adjusting the timings from Kato by –0.0019 d. Table 4 lists the superhump cycle numbers and times of maximum obtained. A weighted linear fit to these gives the ephemeris

$$2454743.1064(8) + 0.07117(1) * E \tag{1}$$

The O-C (observed minus calculated) times relative to this ephemeris are included in Table 4 and plotted in Figure 4. This shows a distinct shortening in the superhump period $P_{sh}$ at about superhump cycle 80 (JD 2454748.8). Mean values of $P_{sh}$ before and after this point are listed in Table 5.

We investigated the possibility of period variation in the cycle range 0 – 80 by applying a 2$^{nd}$ order polynomial fit to these times of maximum. This gives a rate of period change $dP_{sh}/dt = 3.4(2.0) * 10^{-5}$ which we consider a marginal detection. For comparison with previously published values of period derivative for complete outbursts, we also calculated a period derivative based on all of the superhump timings taken together of $dP_{sh}/dt = -9.1(6) * 10^{-5}$, consistent with values seen in other SU UMa systems with a similar $P_{sh}$ [7].

**Light curve period analysis**

A period analysis of our time series data using the Lomb-Scargle method in Peranso [8,9,10] gives the power spectrum shown in Figure 5. The period of the strongest signal is 0.0710(2) d, consistent with the mean $P_{sh}$ value of 0.07117(1) d obtained from the superhump timing analysis. Figure 6 shows the phase diagram obtained by folding the time series data on a period of 0.07117 d. The mean superhump amplitude is 0.12 magnitude.

The period analysis was repeated for each of the above two superhump cycles ranges. Table 5 gives the values of $P_{sh}$ found. These are consistent with the periods from analysing superhump timings. In both cases the superhump signal was removed and the period analysis repeated to look for residual signals. No signals consistent between the two time intervals and above the background noise level remained. In particular, no orbital signal could be resolved.

**Discussion**

Multiple rebrightenings (also known as echo outbursts) are one of the signatures of the WZ Sge sub-class of SU UMa variables [11]. However late rebrightenings, although unusual, have been observed before in normal SU UMa variables [12,13]. The absence of a previous recorded outburst of the new dwarf nova may indicate a long superoutburst recurrence time (supercycle period), another signature of WZ Sge variables, but this outburst was relatively faint and it is possible that previous outbursts were missed. On balance, the modest outburst amplitude and the relatively longer superhump period lead us to believe that this is more likely to be a normal SU UMa dwarf nova than a WZ Sge star. A long supercycle period would indicate a relatively low mass transfer rate.

**Conclusion**

The transient object seen for the first time on 2008 October 3 and observed for a further 24 days has been identified as a SU UMa-type dwarf nova in superoutburst. Its position was measured as 19h 14m 43.52s +60° 52’ 14.1” +/- 0.13” (J2000). The outburst reached magnitude 14.9, its amplitude was approximately 6 magnitudes and it lasted at least 11 days. A brief rebrightening by more than 2 magnitudes was observed about 11 days after the end of the outburst. The superhump amplitude was initially 0.2 magnitude and diminished as the outburst continued. The mean superhump period was 0.07117(1) d but there was a shortening in the period around superhump cycle 80. Before this point we found $P_{sh}$ = 0.07137(2)d and after $P_{sh}$ = 0.07091(2) d. We found weak evidence for an increasing $P_{sh}$ before cycle 80 with $dP_{sh}/dt = 3.4(2.0) * 10^{-5}$.

**Acknowledgements**

We acknowledge with thanks VSNET for its announcement of the discovery and subsequent observations, the AAVSO for variable star observations from its International Database contributed by observers worldwide and used in this research, the NASA Astrophysics Data System, and the Simbad and VizieR Services operated by CDS Strasbourg. We also acknowledge use of the Bradford Robotic Telescope on Tenerife operated by the Department of Cybernetics, University of Bradford. We are grateful for helpful comments by the referee.

**Addresses**

DB: 5 Silver Lane, West Challow, Wantage, Oxon, OX12 9TX, UK [drsboyd@dsl.pipex.com]
KG: 23746 Schoolhouse Road, Manhattan, IL 60442, USA [kag@core.com]
TK: Department of Astronomy, Kyoto University, Kyoto 606-8502, Japan [tkato@kusastro.kyoto-u.ac.jp]
RK: CBA Colorado, 980 Antelope Drive West, Bennett, CO 80102, USA [bob@antelopehillsobservatory.org]
IM: Furzehill House, Ilston, Swansea, SA2 7LE, UK [furzehillobservatory@hotmail.com]
AO: Verkkoniementie 30, FI-40950 Muurame, FINLAND [arto.oksanen@jklsirius.fi]
RP: 3 The Birches, Shobdon, Leominster, Herefordshire, HR6 9NG, UK [roger.pickard@sky.com]
GP: 67 Ellerton Road, Kingstanding, Birmingham B44 0QE, UK [garypoyner@blueyonder.co.uk]

| Start time (JD) | Duration (h) | No obs | Filter | Observer |
|---|---|---|---|---|
| 2454747.398 | 4.29 | 122 | V | RP |
| 2454748.215 | 0.95 | 37 | C | AO |
| 2454748.283 | 5.08 | 268 | C | DB |
| 2454748.367 | 3.38 | 96 | C | RP |
| 2454748.427 | 1.33 | 136 | C | IM |
| 2454748.586 | 4.81 | 228 | C | RK |
| 2454749.280 | 1.00 | 56 | C | DB |
| 2454749.525 | 3.50 | 41 | L | KG |
| 2454750.212 | 2.59 | 91 | C | AO |
| 2454750.503 | 4.04 | 46 | L | KG |
| 2454751.235 | 0.21 | 10 | C | AO |
| 2454751.317 | 0.18 | 20 | C | IM |
| 2454751.413 | 2.63 | 130 | C | DB |
| 2454751.497 | 4.19 | 49 | L | KG |
| 2454752.277 | 0.22 | 25 | C | IM |
| 2454752.281 | 4.52 | 237 | C | DB |
| 2454753.284 | 0.04 | 2 | V | DB |
| 2454761.356 | 0.41 | 38 | C | DB |
| 2454764.463 | 0.02 | 1 | C | GP |
| 2454766.269 | 0.69 | 20 | V | DB |

Table 1. Log of time series observations.

| Observer | Equipment used |
|---|---|
| DB | 0.25-m f/3.6 Newtonian + HX516 CCD |
| KG | 0.30-m f/10 SCT + ST9 CCD |
| RK | 0.25-m f/10 SCT + Apogee AP-47 CCD |
| IM | 0.35-m f/10 SCT + SXVF-H16 CCD |
| AO | 0.40-m f/8.4 Ritchey-Chretien + STL-1001E CCD |
| RP | 0.30-m f/10 SCT + SXV-H9 CCD |
| GP | 0.35-m f/10 SCT + FLI CCD (BRT) |

Table 2. Equipment used.

| USNO-A2.0 identifier | RA (J2000) | Dec (J2000) | V mag |
|---|---|---|---|
| 1500-06789476 | 19 15 12.24 | +60 51 21.9 | 13.31 |
| 1500-06785912 | 19 14 40.68 | +60 49 24.3 | 14.11 |
| 1500-06786144 | 19 14 42.77 | +60 49 49.1 | 14.25 |
| 1500-06783518 | 19 14 19.29 | +60 49 52.7 | 14.86 |
| 1500-06787615 | 19 14 55.78 | +60 52 12.0 | 14.95 |
| 1500-06787821 | 19 14 57.74 | +60 49 46.1 | 15.50 |

Table 3. Comparison stars.

| Superhump cycle no | Observed time of maximum (HJD) | Uncertainty (d) | O-C (d) |
|---|---|---|---|
| 0 | 2454743.0946 | 0.0021 | -0.0118 |
| 1 | 2454743.1678 | 0.0023 | -0.0098 |
| 40 | 2454745.9448 | 0.0013 | -0.0084 |
| 41 | 2454746.0159 | 0.0012 | -0.0084 |
| 61 | 2454747.4489 | 0.0020 | 0.0013 |
| 62 | 2454747.5205 | 0.0029 | 0.0016 |
| 68 | 2454747.9423 | 0.0013 | -0.0036 |
| 69 | 2454748.0164 | 0.0010 | -0.0006 |
| 70 | 2454748.0869 | 0.0011 | -0.0013 |
| 72 | 2454748.2326 | 0.0004 | 0.0021 |
| 73 | 2454748.3036 | 0.0013 | 0.0019 |
| 73 | 2454748.3059 | 0.0012 | 0.0042 |
| 74 | 2454748.3739 | 0.0013 | 0.0011 |
| 74 | 2454748.3734 | 0.0016 | 0.0005 |
| 75 | 2454748.4454 | 0.0018 | 0.0014 |
| 75 | 2454748.4456 | 0.0010 | 0.0016 |
| 75 | 2454748.4456 | 0.0006 | 0.0016 |
| 78 | 2454748.6601 | 0.0010 | 0.0026 |
| 79 | 2454748.7292 | 0.0006 | 0.0005 |
| 82 | 2454748.9472 | 0.0010 | 0.0050 |
| 83 | 2454749.0153 | 0.0015 | 0.0019 |
| 84 | 2454749.0851 | 0.0013 | 0.0005 |
| 87 | 2454749.2972 | 0.0017 | -0.0008 |
| 87 | 2454749.2982 | 0.0019 | 0.0001 |
| 88 | 2454749.3663 | 0.0039 | -0.0029 |
| 91 | 2454749.5844 | 0.0040 | 0.0017 |
| 101 | 2454750.2927 | 0.0009 | -0.0017 |
| 111 | 2454750.9984 | 0.0021 | -0.0077 |
| 112 | 2454751.0614 | 0.0065 | -0.0159 |
| 116 | 2454751.3594 | 0.0014 | -0.0025 |
| 117 | 2454751.4254 | 0.0016 | -0.0076 |
| 117 | 2454751.4319 | 0.0024 | -0.0012 |
| 124 | 2454751.9090 | 0.0029 | -0.0223 |
| 125 | 2454752.0094 | 0.0058 | 0.0070 |
| 126 | 2454752.0645 | 0.0054 | -0.0091 |
| 130 | 2454752.3502 | 0.0024 | -0.0081 |
| 138 | 2454752.9099 | 0.0041 | -0.0177 |
| 139 | 2454752.9929 | 0.0031 | -0.0059 |
| 140 | 2454753.0550 | 0.0030 | -0.0150 |

Table 4. Times of superhump maximum and O-C values relative to the ephemeris in eqn (1).

| Superhump cycle range | HJD range | $P_{sh}$ from superhump timings (d) | $P_{sh}$ from Lomb-Scargle analysis (d) |
|---|---|---|---|
| 0 - 80 | 2454743.0 - 2454748.8 | 0.07137(2) | 0.0711(7) |
| 81 - 140 | 2454748.8 - 2454753.1 | 0.07091(2) | 0.0709(4) |

Table 5. Mean values of $P_{sh}$ before and after superhump cycle 80.

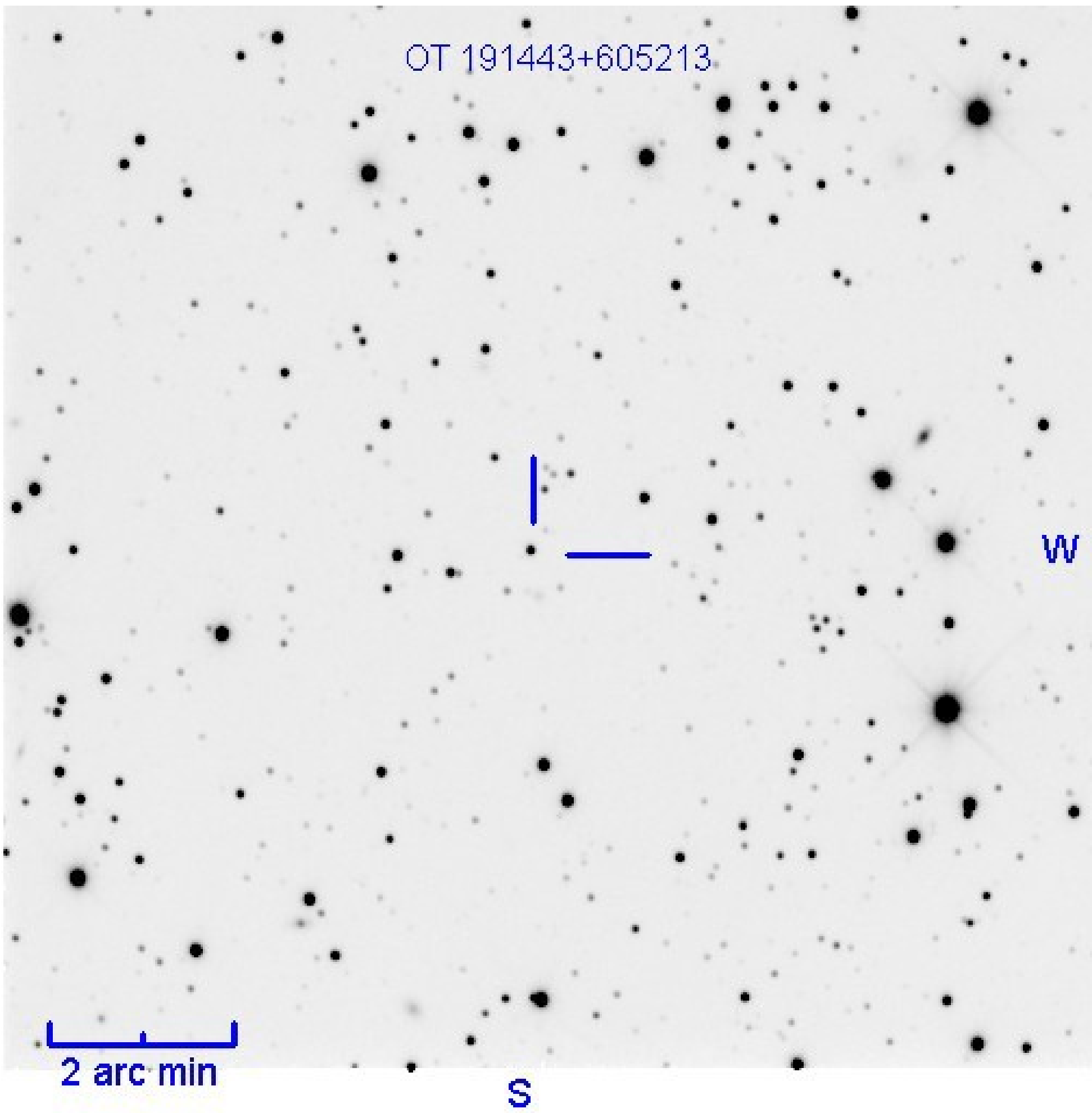

Figure 1. Image of the dwarf nova taken by Oksanen on October 8.

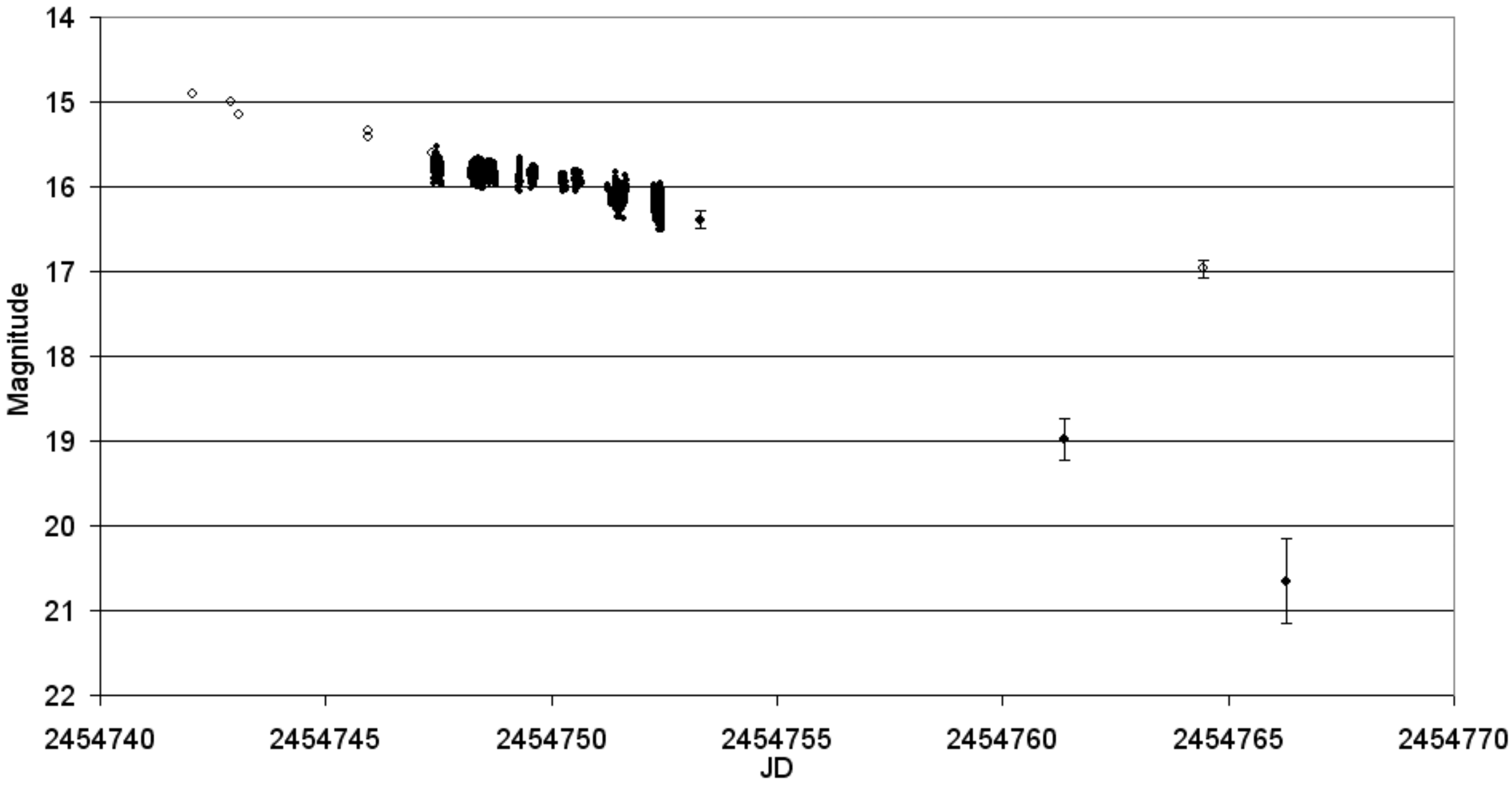

Figure 2. Light curve of the dwarf nova outburst - solid dots without error bars denote time series data, solid dots with error bars are mean values and open dots are single observations.

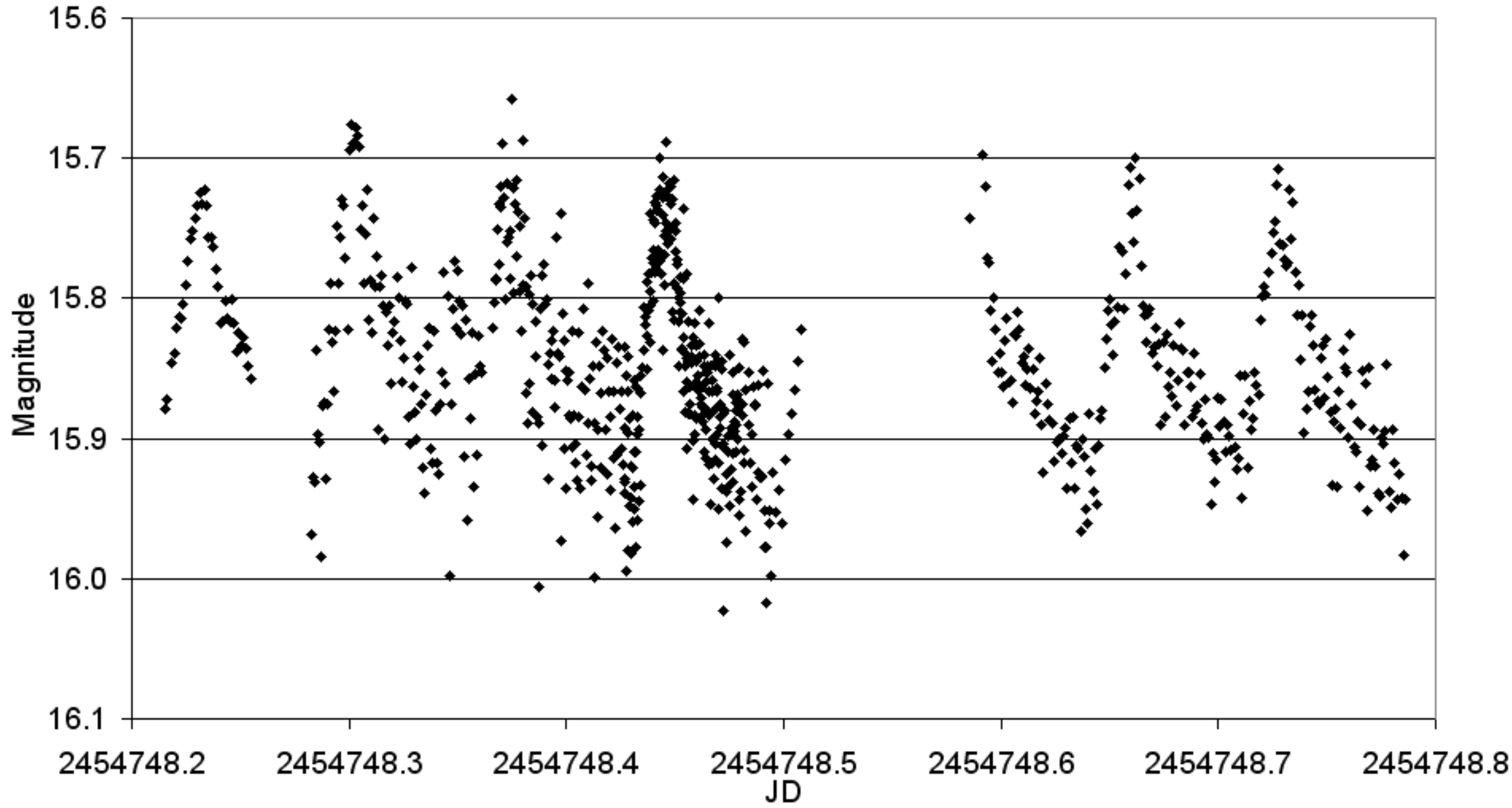

Figure 3. Superhumps on JD 2454748.

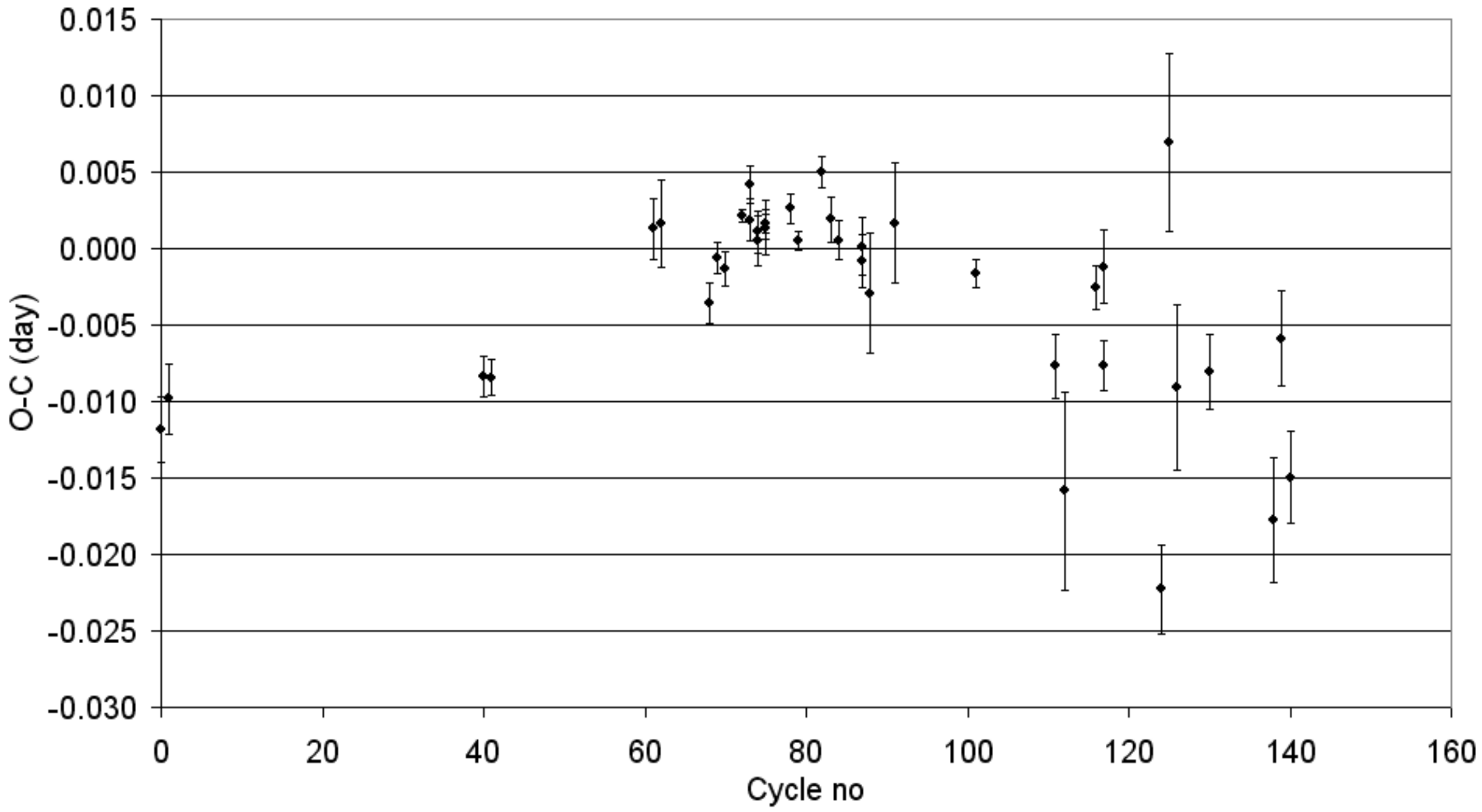

Figure 4. O-C times of superhump maximum with respect to the ephemeris in eqn (1).

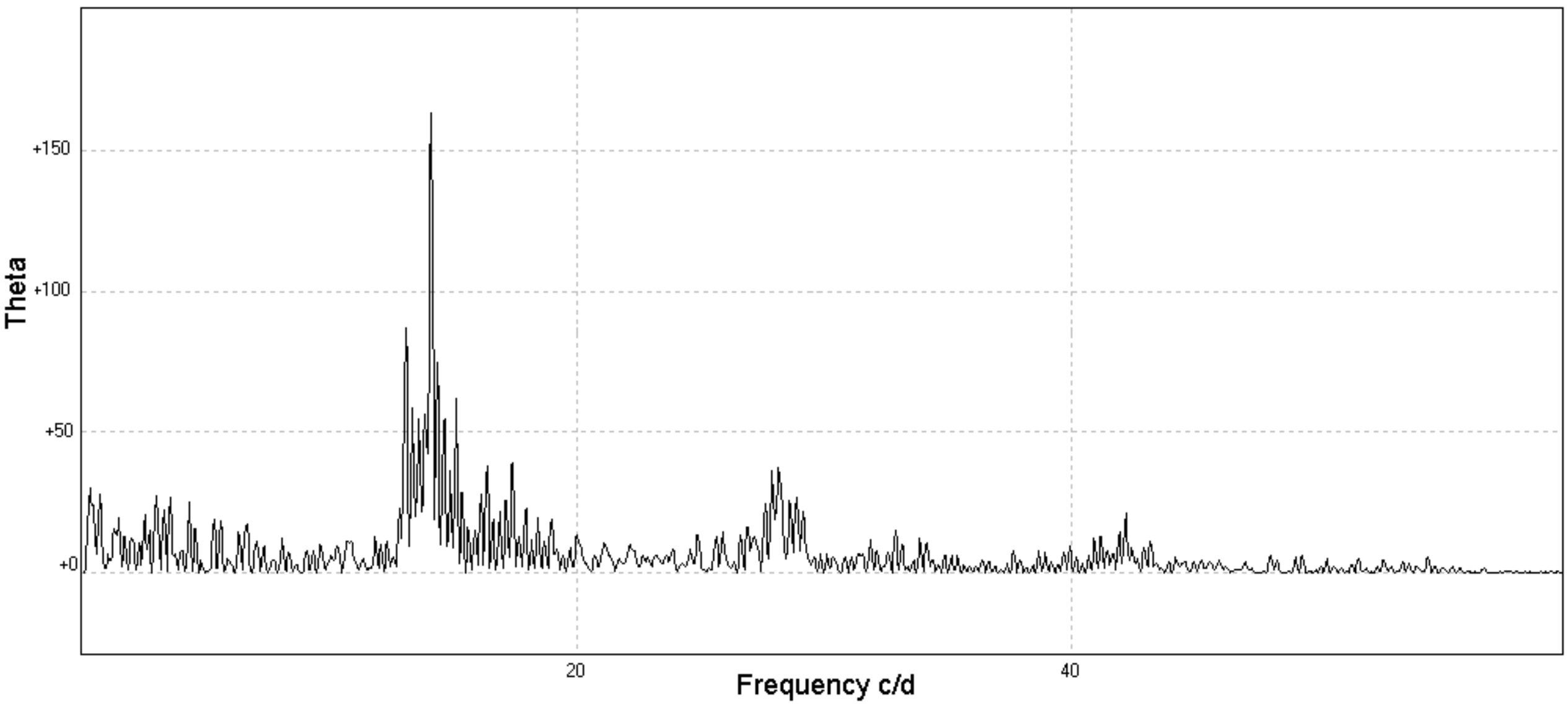

Figure 5. Power spectrum from Lomb-Scargle analysis.

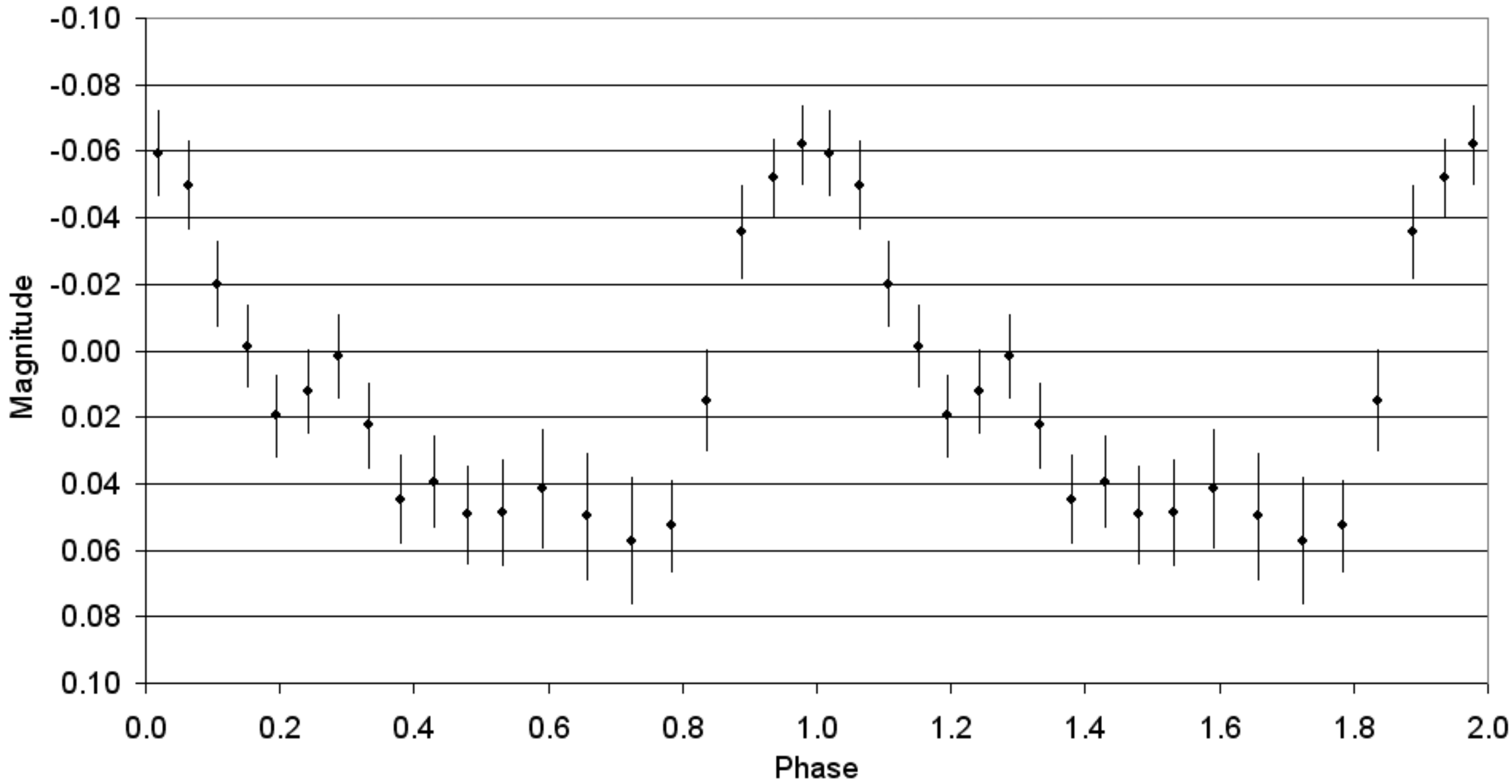

Figure 6. Phase diagram for a superhump period of 0.07117 d (showing two cycles).